\documentclass[draftcls,onecolumn]{IEEEtran}
\IEEEoverridecommandlockouts

\usepackage{amsmath,amssymb,latexsym,cite}
\usepackage{pstricks,graphicx}

\makeatletter
\def\ps@headings{%
\def\@oddhead{\mbox{}\scriptsize\rightmark \hfil \thepage}%
\def\@evenhead{\scriptsize\thepage \hfil \leftmark\mbox{}}%
\def\@oddfoot{}%
\def\@evenfoot{}}
\makeatother

\newtheorem{corollary}{Corollary}

\newtheorem{claim}{Claim}
\newtheorem{theorem}{Theorem}

\def\expe{\mathbb{E}}   %
\def\prob{\mathbb{P}}   %

\newcommand{\norm}[1]{\left\| #1 \right\|}

\def\mc{\mathcal}
\def\mbf{\mathbf}

\def\trel{T_{\mathrm{relax}}}
\def\tave{T_{\mathrm{ave}}}

\newcommand{\ones}{\ensuremath{\vec{1}}}

\newcommand{\beq}{\begin{equation}}
\newcommand{\eeq}{\end{equation}}
\newcommand{\stexp}{\mbox{$\mathbb{E}$}}    %

\usepackage{algorithmic}
\usepackage{url}

\begin{document}

\title{The Impact of Mobility on Gossip Algorithms}

\author{Anand~D.~Sarwate,~\IEEEmembership{Member,~IEEE,}
	and~Alexandros~G.~Dimakis~\IEEEmembership{Member,~IEEE}
\thanks{A. D. Sarwate is with the Information Theory and Applications Center, University of California, San Diego, La Jolla, CA 92093-0447 USA email: \texttt{asarwate@ucsd.edu}}
\thanks{A. G. Dimakis is with the, Department of Electrical Engineering Systems Division, University of Southern California, Los Angeles, CA 90089-2560 USA email: \texttt{dimakis@usc.edu}}
\thanks{Manuscript received September xx, 2009.}
}

\maketitle

\begin{abstract}
The influence of node mobility on the convergence time of averaging gossip algorithms in networks is studied.  It is shown that a small number of fully mobile nodes can yield a significant decrease in convergence time.  A method is developed for deriving lower bounds on the convergence time by merging nodes according to their mobility pattern.  This method is used to show that if the agents have one-dimensional mobility in the same direction the convergence time is improved by at most a constant.  Upper bounds are obtained on the convergence time using techniques from the theory of Markov chains and show that simple models of mobility can dramatically accelerate gossip as long as the mobility paths significantly overlap.  Simulations show that these bounds are still valid for more general mobility models that seem analytically intractable, and further illustrate that different mobility patterns can have significantly different effects on the convergence of distributed algorithms.
\end{abstract}

\section{Introduction}

Gossip algorithms are distributed message passing schemes that are used to disseminate and process information in networks. Average consensus\cite{Tsitsiklis84,Fagn08,MurraySurvey} and 
averaging gossip algorithms~\cite{KempeDG:03gossip,BoydIT} form an important special case of schemes that can compute linear functions of the data in a robust and distributed way. Such schemes have found
numerous uses for distributed estimation, localization and optimization~\cite{SpanosOM:05Kalman,XiaoBL:05ipsn,SaligramaTSP} and also for compressive sensing of sensor measurements and field estimation~\cite{Rabbat:06IPSN}.
In this paper we study gossip algorithms that compute linear functions and will not discuss related problems like information dissemination (see e.g. {\cite{MoskAoyamaS:05gossip,SanghaviIT} and references therein). 

Gossip algorithms are a natural fit for wireless ad-hoc and sensor network applications because of their distributed and robust nature. Recently the broadcast nature of wireless communication has been exploited to improve convergence~\cite{AYSBroadcast07,UstebayOCR:10greedy}.
Another key feature of some wireless networks is node \textit{mobility}; to the best of our knowledge, the impact of mobility on gossip algorithms has not been significantly investigated.  In this paper we attempt to analyze how mobility can (or cannot) help the convergence of gossip algorithms.   For fixed nodes in a random geometric graph or grid (both popular model topologies for large wireless ad-hoc and sensor networks), standard gossip is extremely wasteful in terms of communication requirements; even optimized standard gossip algorithms on a grid converge very slowly, requiring $\Theta(n^2 \log \epsilon^{-1})$ messages~\cite{BoydIT,DimakisSW:08gossip} to compute the average within accuracy $\epsilon$.  Observe that this is of the same order as requiring every node to flood its estimate to all other nodes.  The obvious solution of averaging numbers on a spanning tree and flooding back the average to all the nodes requires only $\Theta(n)$ messages. Clearly, constructing and maintaining a spanning tree in dynamic and ad-hoc networks introduces significant overhead and complexity, but a quadratic number of messages is a high price to pay for fault tolerance.   In this context, what kind of mobility patterns are beneficial and how many mobile agents are needed to boost the convergence speed?  Our results suggest that certain kinds of mobility can, in some cases, significantly accelerate convergence.  This work is a first step to understanding how mobility can impact the convergence of iterative message-passing schemes, at least for the special case of pairwise averaging where the convergence behavior is better understood.

\textbf{Main Results:}   Our first result is that if $m$ nodes have full mobility and the others are fixed in a grid, the convergence time drops to $\Theta(n^2 /m  \log \epsilon^{-1})$. Therefore, even a vanishingly small fraction of mobile nodes can 
change the order of messages required for convergence.  In particular, if any constant fraction of nodes have full mobility, the convergence time drops to $\Theta(n \log \epsilon^{-1})$, the same order as a fully connected graph.  

Our second result is that some mobility patterns might not be beneficial.
We show that even if all the nodes of the network have one dimensional mobility in the same direction (e.g. horizontal), this yields no benefit in the convergence time, up to constants.  Intuitively, this is because the information must still diffuse across the other direction (e.g. vertical).  Finally we show that one dimensional mobility with a randomly selected direction is as good as full mobility. 

In order to obtain these results, we develop a novel method for deriving \emph{lower bounds} on the convergence time of gossip algorithms 
with mobile nodes by merging nodes with similar mobility regions.  This method is based on a characterization of the convergence time of Markov chains in terms of a functional called the Dirichlet form~\cite{AldousFill:mcbook}.
Our upper bounds are derived using the so-called Poincar\'{e} inequality~\cite{Diaconis} 
and the related canonical path method~\cite{Sinclair}; a version of this technique has also been previously used to study gossip algorithms~\cite{BDTVPath}.  Our techniques are fairly general; while we illustrate applications to grid networks and random geometric graphs, the methods can be applied to general graphs.

\section{Network model and preliminaries}

\subsection{Time model}

We use the asynchronous time model \cite{BertsekasBook,BoydIT}, which is well-matched to the distributed nature of wireless networks. In particular, we assume that each sensor has an independent clock whose ``ticks'' are distributed as a rate $\lambda$ Poisson process.  Our analysis is based on measuring time in terms of the number of ticks of an equivalent single
virtual global clock ticking according to a rate $n\lambda$ Poisson
process.  An exact analysis of the time model can be found in \cite{BoydIT}. We will refer to the time between two consecutive clock ticks as one timeslot.  This modeling assumption results in a discrete-time system in which one sensor is selected uniformly in each time slot.

Throughout this paper we will be analyzing the number of required messages without worrying about delay. We can therefore adjust the length of the timeslots relative to the communication time so that only one packet exists in the network at each timeslot with high probability. Note that this assumption is made only for analytical convenience; in a practical implementation, several packets might co-exist in the network, but the associated 
issues are beyond the scope of this work.

\subsection{Network and mobility model}

Suppose we have a collection of $n$ agents $\mc{A}$.  At the first timeslot, each agent $i$ starts at some initial location with a scalar $x_i(0)$. We will denote the vector of their initial values by $\mbf{x}(0)$.  The objective of our algorithm is for every agent to estimate the average
	\begin{align}
	\bar{x} = \frac{1}{n} \sum_{i=1}^{n} x_i(0).
	\end{align}
In order to accomplish this goal, the agents pass messages between each other to communicate their estimates. We assume that this communication always succeeds.
We also assume that the messages involve real numbers; the effects of message quantization in gossip and consensus algorithms is an active area of research \cite{KashyapBS:07quant,Nedic07,Aysal07,NedicOOT:09dist,CarliBZ:10dynamic,CarliFFZ:10quantgossip,KarM:10quant,LavaeiM:11gossip,ZhuM:11time}.

The $n$ agents can move in an area $\mc{G}$.   For example, we may take $\mc{G}$ to be a graph with vertex set $\mc{V}$ and edge set $\mc{E}$.  Agents at locations $v$ and $v'$ can communicate if either $v = v'$ or $(v,v') \in \mc{E}$.  Another example is taking $\mc{G}$ to be the unit square and allowing agents at $v$ and $v'$ to communicate if the distance $d(v,v')$ is less than some radius $r(n)$.  For each location $l$ in $\mc{G}$ there is a set of locations $\mc{N}(l) \subseteq \mc{G}$ such that an agent at $l$ can communication with agents in $\mc{N}(l)$.  If $l' \in \mc{N}(l)$ then $l \in \mc{N}(l')$.

In this paper we will use two networks to illustrate our results.  However, the methods we describe can be used for more general networks with bidirectional communication.
	\begin{enumerate}
	\item Our first example is the $\sqrt{n} \times \sqrt{n}$ discrete lattice on the torus.  The set of locations $\mc{V}$ is $\{0, 2, \ldots, \sqrt{n}-1\}^2$ and there are edges between $(i,j)$ and $(i',j')$ if $i' = (i \pm 1) (\mod \sqrt{n})$ and $j' = (j \pm 1) (\mod \sqrt{n})$.  There are $n$ agents, one for each location in $\mc{V}$, and at time $0$ they each occupy distinct locations in $\mc{V}$.  For a location $(i,j)$ we call the $i$ the row coordinate and $j$ the column coordinate.  
	\item The second example is the random geometric graph (RGG) model on the unit torus.  The unit torus is formed from the unit square by ``glueing'' opposite edges together.  The agent locations are in $[0,1]^2$ and the initial positions of the agents are chosen uniformly in $[0,1]^2$.  Agents can communicate with each other if the distance between them on the torus is less than $r(n) = \sqrt{5 c \frac{\log n}{n}}$, where $c \ge 10$ ensures some useful regularity properties \cite{BDTVPath} discussed subsequently.  Again for an agent at $(i,j)$, we call the $i$ the row coordinate and $j$ the column coordinate.
	\end{enumerate}
	
Under agent-based mobility, at each time step agent $i$ moves to a new location in $\mc{G}$ chosen according to a fixed probability distribution $\mu_i$.  Therefore the sequence of agent locations $l_i(1), l_2(t), \ldots, l_i(t)$ is independent and identically distributed (iid) according to the distribution $\mu_i$.  We call the collection of distributions $\{\mu_i : i \in \mc{A}\}$ an agent-based mobility pattern.  Our theoretical results in this paper are for agent-based mobility.  In particular, we study a few simple examples of mobility.
\begin{enumerate}
\item A simple example of agent-based mobility is \emph{full uniform mobility}.  In this model, $\mu_i$ is the uniform distribution on $\mc{G}$ for each $i \in \mc{A}$.  This corresponds to the case where each agent is equiprobably at any location in the $\mc{G}$ at time $t$.  This is similar to the model proposed by Grossglauser and Tse~\cite{GrossglauserT:02mobility}.  We will also consider a static network with $m$ fully mobile agents added to the network.
\item In the \textit{horizontal mobility} model, each agent selects a new horizontal location uniformly at each time.  For the torus, the agent selects a new column coordindate uniformly from $\{1, 2, \ldots, \sqrt{n}\}$.  For the RGG, it selects a new horizontal coordinate uniformly from $[0,1]$.  
\item In the \textit{bidirectional} model each agent selects equiprobably whether it will move horizontally or vertically for all time.  At each time step, the horizontal agents select a new horizontal coordinate uniformly, and the vertical agents select a new vertical coordinate uniformly.
\item In a \textit{local} model for the torus, an agent that starts initially at location $(i,j)$ chooses a new location uniformly in the square of size $(2m+1)^2$ centered at $(i,j)$.  That is, the  horizontal coordinate is uniformly distributed in $\{i-m , \ldots, i+m \} \mod \sqrt{n} $ and the vertical coordinate is chosen uniformly in $\{j-m, \ldots, j+m \} \mod \sqrt{n}$.  Once the new coordinates are chosen, an agent can communicate with other agents in the same or adjacent locations in the $\sqrt{n} \times \sqrt{n}$ torus.
\end{enumerate}

The key assumption in all our mobility models is that in each gossip timeslot, the positions of the mobile agents are selected \emph{independently} from some distribution supported on a sub-region of the space, similarly to Grossglauser and Tse \cite{GrossglauserT:02mobility}.
Popular mobility models like the random walk model~\cite{GrossglauserV:06ease,latticeRW}, random waypoint model~\cite{RandWaypoint98}, and random direction model~\cite{RandDirection01} have time dependencies. If however the duration of one gossip timeslot is comparable or larger than the mixing time of the mobility model, the positions of the agents will be approximately independent (see also~\cite{Spyropoulos08}). If delay is not an issue, we can always set the duration of the gossip timeslot to have that property, and in simulations we show that if we do not allow the mobility model to mix, mobility is not as helpful.  We believe that  our analytic results could be used to bound these more realistic mobility models, but we leave this for future work.

\section{Algorithm and main results}

\subsection{The algorithm}

The gossip algorithm that we will consider is a simple extension of the standard 
nearest-neighbor gossip of Boyd et al. \cite{BoydIT} that includes the mobility model in a natural way.  At each time step, the agents move independently to new locations. One agent is selected at random, chooses one of its neighbors according to the graph $G$, and performs a pairwise average with that neighbor.  More precisely, at each time $t = 1, 2, \ldots$ the following events occur:

\begin{enumerate}
\item Each agent $i \in \mc{A}$ chooses a new location $l_i(t)$ according to the mobility distribution $\mu_i$.
\item A agent $i$ is selected at random and selects a neighbor $j$ uniformly from the set $\mc{N}(l_i(t))$.  For example, if $\mc{G}$ is a graph, then
	\begin{align}
	\mc{N}(l_i(t)) = \{k  \in \mc{V} :  (l_i(t),l_k(t)) \in \mc{E} \}.
	\end{align}
\item The agents $i$ and $j$ exchange values and update their estimates:
	\begin{align}
	x_k(t) = \left\{
		\begin{array}{ll}
		\frac{1}{2} (x_i(t-1) + x_j(t-1))   &   k = i, j \\
		x_k(t-1)                            &   k \ne i, j
		\end{array}
		\right.
	\end{align}
\end{enumerate}

Since the algorithm is randomized, we are interested in providing probabilistic bounds on its running time.   Given $\epsilon > 0$, the $\epsilon$-averaging time~\cite{BoydIT} is the earliest
time at which the vector $\mbf{x}(t)$ is $\epsilon$ close to the normalized
true average with probability greater than $1 - \epsilon$: 
	\begin{align} 
	\tave(n,\epsilon) & 		\nonumber \\
		&\hspace{-0.8cm}
		= \sup_{x(0)} \inf_{t = 0,1,2 \ldots}
			\left\{ \prob\left( 
				\frac{ 
					\norm{\mbf{x}(t)- \bar{x} \ones} 
					}{ 
					\norm{\mbf{x}(0)} 
					} 
				\geq
				\epsilon \right) 
			\leq \epsilon \right\},
	\label{ave_time_definition}
	\end{align}
where $\norm{\cdot}$ denotes the Euclidean norm.  Note that this is
essentially measuring a rate of convergence in probability.
The analysis of Denantes et al.~\cite{DenantesBenezit} shows that bounds on the spectral gap yield an asymptotic deterministic rate of vanishing error.  Our bounds can be used to bound both the rate of convergence in probability and to show that the averaging error decays exponentially asymptotically almost surely.

\subsection{Main results}

Our main results characterize the benefit (or lack thereof) of mobility in speeding up the convergence of gossip algorithms.  For the network on the grid or torus with no mobility, the averaging time is $\tave^{(\mathrm{torus,none})}(n,\epsilon) = \Theta(n^2 \log \epsilon^{-1})$.  For the network on the random geometric graph with the connectivity radius chosen as described above, The averaging time is $\tave^{(\mathrm{RGG,none})}(n,\epsilon) = \Theta( \frac{n^2}{\log n} \log \epsilon^{-1})$.
\begin{itemize}
\item For horizontal mobility on the random geometric graph and the torus, the averaging time improves by at best a constant factor over the case where the agents are not mobile at all:
	\begin{align}
	\tave^{(\mathrm{torus,horiz})}(n,\epsilon) &= \Omega( n^2 \log \epsilon^{-1} )  \\
	\tave^{(\mathrm{RGG,horiz})}(n,\epsilon) &= \Omega \left( \frac{n^2 \log \epsilon^{-1}}{\log n} \right) .
	\end{align}
\item For bidirectional mobility where each agent initially selects whether to move vertically or horizontally, the convergence time is within a constant factor of full mobility:
	\begin{align}
	\tave^{(\mathrm{torus,bi})}(n,\epsilon) &= O\left( n \log \epsilon^{-1} \right) \\
	\tave^{(\mathrm{RGG,bi})}(n,\epsilon) &= O\left( n \log \epsilon^{-1} \right).
	\end{align}
\item For $n$ non-mobile agents on a $\sqrt{n}\times\sqrt{n}$ torus with $m \le n$ agents having full mobility, the convergence time is
	\begin{align}
	\tave^{(\mathrm{torus\ plus\ m,2D})}(n,\epsilon) &= \Theta\left( \frac{n^2}{m} \log \epsilon^{-1} \right).
	\end{align}
\item For the local mobility model with each agent moving in a square of size $(2m+1)^2$, 
	\begin{align}
	\tave^{(\mathrm{torus,local})}(n,\epsilon) &= O\left( \frac{n^2 \log m}{m^2} \log \epsilon^{-1} \right).
	\end{align}
\end{itemize}

\section{Upper and lower bounds on convergence time}

\subsection{Convergence analysis}

At each step of the algorithm, the agents update their estimates of the average $\bar{x}$.  Let $\mbf{x}(t)$ denote the average estimates at time $t$.  For agents $i$ and $j$ define the matrix $W^{(i,j)}$
	\begin{align}
	W^{(i,j)} = I - \frac{1}{2} (\mbf{e}_i - \mbf{e}_j) (\mbf{e}_i - \mbf{e}_j)^T,
	\end{align}
where $\mbf{e}_i$ is the vector with $1$ in the $i$-th coordinate and $0$'s elsewhere.  If the pair $(i,j)$ average at time $t$ then new vector of averages is given by
	\begin{align}
	\mbf{x}(t) = W^{(i,j)} \mbf{x}(t-1).
	\end{align}
The randomness in the mobility and in the agent selection induces a probability distribution on the matrices $\{W^{(i,j)} : i, j \in \mc{A}\}$.  Since the mobility and selection are iid across time, we can write the update as
	\begin{align}
	\mbf{x}(t) = \left( \prod_{s=1}^{t} W(s) \right) \mbf{x}(0),
	\end{align}
where $\{W(s)\}$ are iid random matrices.  Denote the expected value of this random matrix by $\bar{W} = \expe[W(s)]$.  It is not hard to see that $\bar{W}$ is a (symmetric) stochastic matrix and therefore corresponds to a Markov chain.  Let $P_{ij}$ be the probability that agent $i$ is selected in step 2 of the algorithm and it selects agent $j$ in its neighbor set.  Then it is clear that $\prob(W(s) = W^{(i,j)}) = P_{ij} + P_{ji}$, and that
	\begin{align}
	\bar{W}_{ij} = \frac{1}{2} ( P_{ij} + P_{ji} ).
	\label{eq:W_entries}
	\end{align}

The pioneering work of Boyd et al. \cite{BoydIT} showed that the convergence time of a randomized gossip algorithm is dictated by the mixing time of the Markov chain associated to $\bar{W}$.  Mathematically, our problem is how to analyze the mixing time of the new graph induced by the new feature (in this case mobility) and then compare it to the old graph without mobility.  For a Markov chain $\mc{M}$ with transition matrix $\bar{W}$, the convergence rate to the stationary distribution is given by $\lambda_2(\bar{W})$, the second largest eigenvalue of $\bar{W}$. Note that the largest eigenvalue $\lambda_1(\bar{W})$ is $1$.  Define the relaxation time $\trel$ to be the reciprocal of the spectral gap:
	\begin{align}
	\trel( \bar{W} ) = \frac{1}{1 - \lambda_2(\bar{W})}.
	\end{align}
The following theorem is implicit in \cite{BoydIT} (see also \cite{DimakisKMRS:10survey}).

\begin{theorem}[Convergence with $\trel$ \cite{BoydIT,DimakisKMRS:10survey}]
\label{thm:boydthm}
If $P = (P_{ij})$ is symmetric and $n$ is sufficiently large, then $T_{\mathrm{ave}}(n,\epsilon)$ is bounded by
	\begin{align}
	\tave(n,\epsilon) = \Theta \left( \trel(\bar{W}) \log \epsilon^{-1} \right)
	\end{align}
\end{theorem}

\subsection{Lower bounds}

In this section we provide a general method for constructing lower bounds on the convergence time for pairwise gossip algorithms under agent-based mobility.  The main intuition is to partition the set of vertices in the graph and merge all agents whose mobility is supported in the same element of the partition.  This induces a transformation on the Markov chain associated to the gossip algorithm.  By using an extremal characterization of the relaxation time for Markov chains we can lower bound the $\trel(\bar{W})$ in the original gossip algorithm by that for the induced Markov chain.  The only remaining issue is to choose a partition that yields a tight lower bound.  At the moment, this must be done by inspection, but we can use this technique to show that horizontal mobility cannot improve the convergence of gossip for the torus or the RGG.

\begin{theorem}		\label{thm:lowerbound}
Let $\{\mc{U}_r\}$ be any partition of the set of locations $\mc{G}$, and let $\hat{W}$ be the transition matrix of the chain induced by merging all agents whose mobility is restricted to a single set in the partition.  Then
	\begin{align}
	\tave(n,\epsilon) = \Omega( \trel(\hat{W}) \log \epsilon^{-1} ).
	\end{align}
\end{theorem}

\begin{proof}
We begin with the set $\mc{G}$ on which the agents in $\mc{A}$ can move.   Let $\{ \mc{U}_r : r = 1, 2, \ldots, M\}$ be a partition of $\mc{G}$.  Given an agent-based mobility pattern $\{\mu_i\}$, let
	\begin{align}
	\mc{C}_r = \{ v \in \mc{A} : \mu_v( \mc{U}_r ) = 1 \},
	\end{align}
be the set of agents whose mobility is restricted to $\mc{U}_r$.  We can create a map $F$ on the state set $\mc{A}$ of the Markov chain corresponding to the gossip algorithm:
	\begin{align}
	F(a) = \left\{
		\begin{array}{ll}
		r & \mathrm{if\ } a \in \mc{C}_r \\
		a & \mathrm{otherwise}
		\end{array}
	\right.
	\end{align}
The map $F$ merges agents whose mobility is restricted to $\mc{U}_r$ and leaves the other agents invariant.  Let $\mc{B}$ denote the image of $F$.  For a Markov chain on $\mc{A}$ with transition probabilities $W_{ij}$ and stationary distribution $\pi(\cdot)$, we can define a new Markov chain on $\mc{B}$ with transitions $\hat{W}_{kl}$:
	\begin{align}
	\hat{W}_{kl} = \frac{1}{ \sum_{i : F(i) = k} \pi(i) } \sum_{i : F(i) = k} \sum_{j : F(j) = l} \pi(i) W_{ij}.
	\label{eq:induced_trans}
	\end{align}
This is the \textit{induced chain} from the function $F$ \cite[Chapter 4, p.37]{AldousFill:mcbook}.  The stationary distribution of this chain is $\hat{\pi}(k) = \sum_{i : F(i) = k} \pi(i)$.

We can express the relaxation time of a Markov chain in terms of the Dirichlet form \cite{AldousFill:mcbook}.  Given a real-valued function $g$ on the state space of the Markov chain with transition matrix $W$ and stationary distribution $\pi(\cdot)$, the Dirichlet form is given by
	\begin{align}
	\mc{D}(g,g)
		=
		\frac{1}{2} \sum_{k,l} \pi(k) W_{kl} 
			( g(k) - g(l) )^2.
	\end{align}
The relaxation time is then given by
	\begin{align}
	\trel( W ) = \sup_{g} \left\{
		\frac{ \sum_{k} \pi(k) g(k)^2 
			}{
			\mc{D}(g,g) }
		: \sum_{k} \pi(k) g(k) = 0
		\right\}.
	\label{eq:trel_extremal}
	\end{align}
The following contraction principle shows that $\trel$ for an induced chain is at most that of the original chain. The validity of this claim is mentioned in \cite[Chapter 4, p.37]{AldousFill:mcbook} and here we present a proof which easily follows from similar arguments from~\cite{AldousFill:mcbook}.

\begin{claim}
Let $\mc{M}$ be a Markov chain on a finite state space $\mc{A}$ with transition matrix $W$ and let $F : \mc{A} \to \mc{B}$ be an arbitrary mapping.  Then the relaxation time of the chain $\hat{\mc{M}}$ on $\mc{B}$ with transition matrix $\hat{W}$ given by (\ref{eq:induced_trans}) induced by $F$ lower bounds the relaxation time of the original chain:
	\begin{align}
	\trel( \hat{W} ) \le \trel( W ).
	\end{align}
\end{claim}

We use the extremal property of the relaxation time in (\ref{eq:trel_extremal}).  Let $\hat{g}$ achieve the supremum in (\ref{eq:trel_extremal}) for the induced chain given by $\hat{W}$.
We can create a function $g$ from $\hat{g}$ to lower bound $\trel( \mc{M} )$.  Let $\mc{U}_k = \{i : F(i) = k\}$ for each $k \in \mc{B}$.  Simply set $g(i) = \hat{g}(k)$ for $i \in \mc{U}_k$.  Then 
	\begin{align}
	\sum_{i \in \mc{A}} \pi(i) g(i)^2 = \sum_{k \in \mc{B}} \hat{\pi}(k) \hat{g}(k)^2.
	\end{align}
Note that $\{\mc{U}_k : k \in \mc{B}\}$ forms a disjoint partition of $\mc{A}$.  For this function $g$, using \eqref{eq:induced_trans} yields
	\begin{align*}
	\mc{D}(g,g) &= \frac{1}{2} \sum_{i,j \in \mc{A}} \pi(i) W_{i,j} (g(i) - g(j))^2 \\
	&= \frac{1}{2} \sum_{k,l \in \mc{B}}
		\left( \sum_{i \in \mc{U}_k} \sum_{j \in \mc{U}_l} \pi(i) W_{ij} \right) (\hat{g}(k) - \hat{g}(l))^2 \\
	&= \frac{1}{2} \sum_{k,l \in \mc{B}} \hat{\pi}(k) \hat{W}_{kj},
	\end{align*}
and therefore the Dirichlet form $\mc{D}(g,g) = \mc{D}(\hat{g},\hat{g})$.  Therefore the supremum of (\ref{eq:trel_extremal}) for the original chain is at least as large as that for the induced chain.
\end{proof}

Note that while the mixing time of a Markov chain decreases when states are merged, 
as argued, the same is not true for other quantities like the expected time to go from 
one state to another.  The preceding lemma and Theorem \ref{thm:boydthm} gives a lower bound on the benefit on the convergence speed of gossip in a network of mobile nodes.   In theory we could optimize the lower bound over all partitions $\{\mc{U}_r\}$, but for our examples there is an ``obvious'' partition that yields a meaningful lower bound.  We turn first to the $\sqrt{n} \times \sqrt{n}$ torus.

\begin{corollary}[Torus with horizontal mobility]
Let $G = (\mc{V},\mc{E})$ be the $\sqrt{n} \times \sqrt{n}$ torus and suppose that the set of agents $\mc{A} = \mc{V}$.  Let the mobility pattern for the $(i,j)$-th agent be uniformly distributed on the set $\mc{U}_i \{(i,k) : k \le \sqrt{n}\}$, which corresponds to mobility only in the horizontal direction.  Then 
	\begin{align}
	\tave( n, \epsilon ) 
		= 
		\Omega \left( n^2 \log \epsilon^{-1} \right).
	\end{align}
\label{cor:torusbad}
\end{corollary}

\begin{proof}
Let $\mc{U}_i = \{(i,j) : j = 1, 2, \ldots, \sqrt{n}\}$ be the $i$-th row of the torus, so $\{\mc{U}_i\}$ partitions $\mc{V}$.  Consider two agents, one starting at $(i,j)$ and the other at $(k,l)$, where $k = i \pm 1 \mod \sqrt{n}$.  Then the probability in the algorithm that $(i,j)$ and $(k,l)$ average with each other is the chance that $(i,j)$ is selected times the probability (over the mobility) that $(i,j)$ and $(k,l)$ are adjacent to each other times the chance that $(i,j)$ selects $(k,l)$ out of its neighbors.  We can upper bound this probability:
	\begin{align}
	W_{ij} = O\left( \frac{1}{n} \times \frac{1}{\sqrt{n}} \right).
	\end{align}

The chain induced from this partition is a cycle with $\sqrt{n}$ states, where each state corresponds to a row in the original Markov chain.  The transitions from row to row are given by (\ref{eq:induced_trans}):
	\begin{align}
	\hat{W}_{kl} &= \frac{1}{ \sum_{i : F(i) = k} \pi(i) } \sum_{i : F(i) = k} \sum_{j : F(j) = l} \pi(i) W_{ij} \\
	&= \sqrt{n} \cdot \sqrt{n} \cdot \sqrt{n} \cdot \frac{1}{n} \cdot
		O\left( \frac{1}{n} \times \frac{1}{\sqrt{n}} \right) \\
	&= O\left( \frac{1}{n} \right).
	\end{align}
Therefore the self-transition for each state is $1 - O(1/n)$.  Let $\alpha = \hat{W}_{kl}$, the transitions from row to row.  The matrix $\hat{W}$ is circulant and generated by the vector $(\alpha, 1 - 2 \alpha, \alpha, 0, \ldots, 0)$.  The eigenvalues are given by the discrete Fourier transform of the vector (c.f. \cite{DimakisSW:08gossip}):
	\begin{align}
	\lambda_k(\hat{W}) = 1 - 2 \alpha + 2 \alpha \cos \left( \frac{(k-1) 2 \pi }{\sqrt{n}} \right).
	\end{align}
In particular, the second-largest eigenvalue can be bounded using the Taylor expansion of the cosine:
	\begin{align}
	\lambda_2(\hat{W}) &\ge 1 - 2 \alpha + 2 \alpha \left( 1 - \frac{1}{2} \frac{4 \pi^2}{n} \right) %
	= 1 - O \left( \frac{1}{n^2} \right). \nonumber
	\end{align}
Therefore the relaxation time is
	\begin{align}
	\trel = \Omega( n^{2} ),
	\end{align}
and the averaging time is bounded by Theorem \ref{thm:boydthm}.%
\end{proof}

The preceding theorem shows that allowing nodes to move in only one direction gives the same order convergence time as the the torus without any node mobility.  That is, \textit{sometimes mobility can yield no significant benefits in terms of convergence}.  In the case where we add a single agent moving in the vertical direction we still do not gain anything.  The proof follows from the same arguments as Corollary \ref{cor:torusbad}.

\begin{corollary}[A single vertical mover doesn't help]
Let $G = (\mc{V},\mc{E})$ be the $\sqrt{n} \times \sqrt{n}$ torus and suppose that the set of agents $\mc{A} = \mc{V} \cup \{e\}$.  Let the mobility pattern for the $(i,j)$-th agent in $\mc{V}$ be uniformly distributed on the set $\{(i,k) : k \le \sqrt{n}\}$, which corresponds to mobility only in the horizontal direction.  Let the mobility pattern for $e$ be uniform on $\{ (i,1) : i \in \sqrt{n} \}$.  Then for this gossip algorithm,
	\begin{align}
	\tave( n, \epsilon ) 
		= 
		\Omega \left( n^2 \log \epsilon^{-1} \right).
	\end{align}
\end{corollary}

We could prove in a similar way that adding a constant number of agents in the vertical direction does not speed up the convergence appreciably.  Our final result in this section shows that 1D unidirectional mobility cannot help speed up the convergence time of gossip on random geometric graphs as well.  Boyd et al. \cite{BoydIT} have shown that the averaging time for standard pairwise gossip on the RGG is $\Theta( n r^{-2} \log \epsilon^{-1})$, which for $r(n) = \Theta(\sqrt{n^{-1} \log n})$ is $\Theta( (n^2/\log n) \log \epsilon^{-1})$.

\subsection{Upper bounds}

For our upper bounds we use the canonical path method \cite{Sinclair}, which we summarize here for completeness.
For any ergodic and reversible Markov chain  on a state space $\Omega$, 
for each pair $i,j$ of states define the \emph{capacity} of a directed edge $e = (i,j)$ to be
	\begin{align} 
	C(e)= \pi(i) \bar{W}_{ij}.
	\end{align}
For each pair of states we define a \emph{demand} $D(i,j)=\pi(i) \pi(j)$. 
A \emph{flow} is any way of routing $D(i,j)$ units of ``liquid'' from $i$ to $j$ 
for all pairs $i,j$ simultaneously. Formally, a flow $F:\mathcal{P} \rightarrow \mathbb{R}^{+}$ is a function on the set $\mathcal{P}$ of all simple paths on the transition graph of the Markov chain that satisfies the demand:
	\begin{align}
	\sum_{p \in \mathcal{P}_{ij}} F(p)= D(i,j),
	\end{align}
where $\mathcal{P}_{ij}$ denotes all the paths from $i$ to $j$. 

For a flow $F$ we can define the \emph{load} on an edge $e$ to be total flow routed across that edge:
	\begin{align}
	f(e) = \sum_{i,j \in \Omega} \sum_{p \in \mathcal{P}_{ij} : e \in p} F(p)
	\end{align}
The \emph{cost} of a flow $F$ is the maximum overload of any edge: 
	\begin{align}
	\rho(F) = \max_e \frac{f(e)}{C(e)},
	\end{align} 
Finally, define the \emph{length} of a flow $l(f)$ to be longest flow-carrying path, i.e. the longest $p$ for which $F(p) \ne 0$.  

Using these definitions, we can use the following Poincar\'{e} inequality~\cite{Sinclair} to yield an upper bound on the inverse spectral gap (relaxation time) of the Markov chain: 
	\begin{align}
	\frac{1}{1-\lambda_2(\bar{W})} \leq \rho(F) \l(F).
	\end{align}
Intuitively, if there are no 'bottlenecks' on the transitions for every pair of states, the relaxation time of the chain will be very small.   Any flow $F$ gives an upper bound that depends on the cost $\rho(F)$ of its most congested edge.

\begin{corollary}[Full mobility is optimal]
Let the area in which the agents move be given by the graph $G = (\mc{V},\mc{E})$ corresponding to the $\sqrt{n} \times \sqrt{n}$ discrete lattice on the torus.  Let the set of agents $\mc{A} = \{1, 2, \ldots, \sqrt{n}\}^2$ with initial locations equal to $\mc{V}$.  Suppose the mobility pattern of every agent in $\mc{A}$ is the uniform distribution on the set of all locations $\mc{V}$, which corresponds to full mobility.  Then for this gossip algorithm,
	\begin{align}
	\tave( n, \epsilon ) 
		= 
		\Omega \left( n \log \epsilon^{-1} \right).
	\end{align}
\end{corollary}

\begin{proof}
The stationary distribution is uniform, so $\pi(i) = 1/n$ for all $i$ and the demand $D(i,j) = 1/n^2$ for all pairs $(i,j)$.  Furthermore, the probability of $i$ and $j$ averaging is $\Omega(1/n^2)$, so the state diagram of the Markov chain is the complete graph with edge capacities $\Omega(1/n^3)$.  The simplest flow is to route directly the demand $1/n^2$ on the edge from $i$ to $j$, which gives a cost of $O(n)$ with a flow of length $1$, so the relaxation time is $O(n)$.
\end{proof}

A slightly less simple example is a cycle with one fully mobile agent.  The cycle has averaging time $\Theta(n^3 \log \epsilon^{-1})$ (see \cite{DimakisSW:08gossip}).  With one mobile agent the averaging time drops to $O(n^2 \log \epsilon^{-1})$

\begin{corollary}[Cycle with one fully mobile agent]
Let the area in which the agents move be given by the graph $G = (\mc{V},\mc{E})$ corresponding to the the cycle of length $n$ and let there be $n+1$ agents $\mc{A} = \mc{B} \cup \{v'\}$, where $\mc{B} = \mc{V} = \{1, 2, \ldots n\}$.  The initial locations of the agents in $\mc{B}$ are the locations of $\mc{V}$ and the agents in $\mc{B}$ cannot move.  The agent $v'$ has mobility uniformly distributed on $\mc{V}$ with initial location $1$.  Then for this gossip algorithm,
	\begin{align}
	\tave( n, \epsilon ) 
		= 
		\Omega \left( n^2 \log \epsilon^{-1} \right).
	\end{align}
\end{corollary}

\begin{proof}
The stationary distribution for this chain is uniform, so $\pi(i) = 1/(n+1)$ for all $i$ in $\mc{A}$.  The probability that $i$ and $j$ average for $i,j \in \mc{V}$ is $0$ unless $i$ and $j$ are neighbors.  Otherwise, with probability $\frac{3}{n}$ the mobile node $v'$ is a neighbor of $i$, so:
	\begin{align*}
	P_{ij} = \frac{1}{n} \left( \left(1 - \frac{3}{n}\right) \cdot \frac{1}{2} + \frac{3}{n} \cdot \frac{1}{3} \right) = \frac{1}{2n} \left( 1 - \frac{1}{n} \right).
	\end{align*}
For $i \in \mc{A}$ and $j = v'$ we have
	\begin{align*}
	P_{iv'} = \frac{1}{n} \cdot \frac{3}{n} \cdot \frac{1}{3} = \frac{1}{n^2}.
	\end{align*}
Thus the capacities are
	\begin{align}
	C(i,j) = \left\{ \begin{array}{ll}
		\frac{1}{2n(n+1)} \left( 1 - \frac{1}{n} \right) & j \in \mc{V} \\
		\frac{1}{n^2 (n+1)} & j = v'
		\end{array}
		\right.
	\end{align}
The demand is just $D(i,j) = 1/(n+1)^2$ between each pair of nodes.

To construct a flow $F$, we just route all flow through the mobile agent $v'$.  An edge $(i,v')$ for $i \in \mc{B}$ carries $n$ flows to all agents $j \ne i$, each of size $1/(n+1)^2$ for a total of $f(i,v') = n/(n+1)^2$.  Similarly, any edge $(v',i)$ carries the same total flow.  All flows are of length $2$, so $l(F) = 2$.  The overload is
	\begin{align*}
	\rho(F) = \frac{n/(n+1)^2}{1/(n^2 (n+1))} = \frac{n^3}{(n+1)}.
	\end{align*}
And thus for large $n$ we get an upper bound of $O(n^2)$ for the relaxation time of the chain.  The averaging time then follows from Theorem \ref{thm:boydthm}.
\end{proof}

\section{Examples revisited}

We now turn to our examples of mobility and derive scaling results for gossip with mobility.  For the torus we will show that local mobility in a square of area $m^2$ cuts the convergence time by $m^2$ and adding $m$ fully mobile agents cuts the convergence time by $m$.  For the random geometric graph we will prove the same result for bidirectional mobility and a lower bound for unidirectional mobility.

\subsection{The torus}

\subsubsection{Local mobility}

An important step in bridging the mobility model here with more reasonable mobility models is to consider local mobility, in which an agent moves uniformly in a square of side length $(2m+1)$ centered at its initial location.

\begin{figure}[t]
\begin{center}
 \includegraphics[width=8cm]{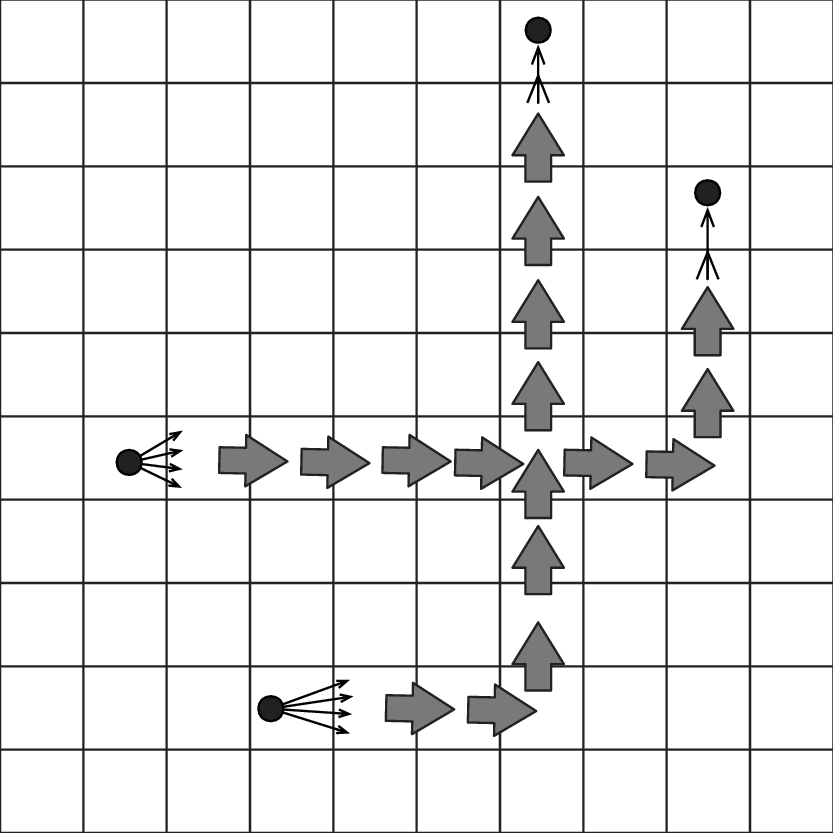}
  \caption{Routing flow in the local mobility model.  Nodes route flows along L-shapde paths through the squares.}
  \label{local_example}
  \end{center}
\end{figure}

\begin{figure}[t]
\begin{center}
 \includegraphics[width=8cm]{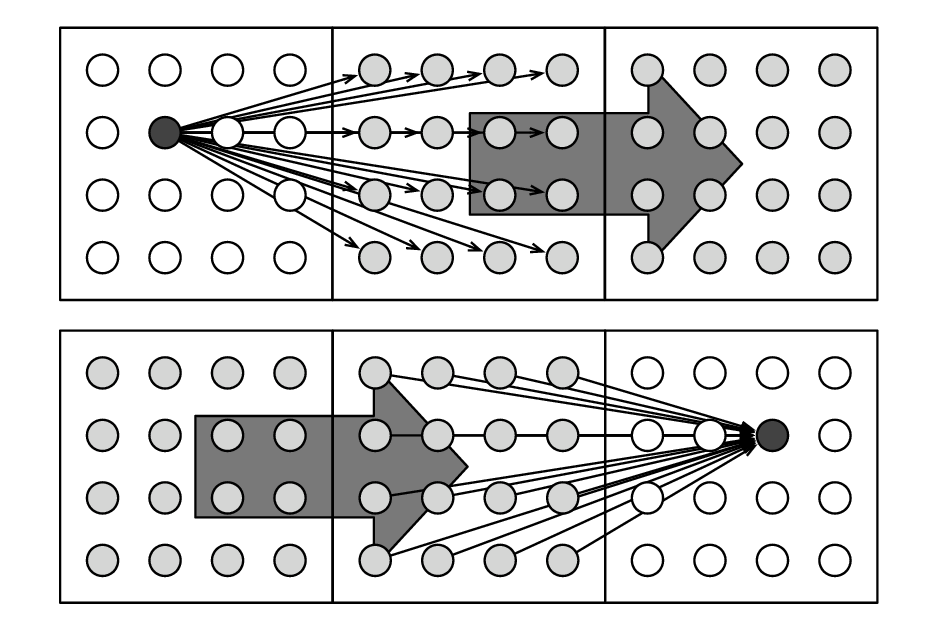}
  \caption{Routing flow in the local mobility model.  As illustrated on top, for node $i$ to send to node $i$, it evenly divides the flow and sends it to all node in the adjacent square in the L-shaped path.  Each node in the adjacent square routes that flow uniformly to every node in the next square in the path.  At the end of the route, as illustrated on the bottom, the nodes in the square adjacent to the destination $j$ transmit their received flows from $i$ directly to $j$.}
  \label{local_example2}
  \end{center}
\end{figure}

\begin{theorem}
Consider gossip with $n$ agents on the $\sqrt{n} \times \sqrt{n}$ torus $\mc{G}$.  Let the agent initially at a location $i$ have mobility uniform in a square of side-length $2m + 1$ centered at $i$.  Then the averaging time is given by
	\begin{align}
	\tave(n,\epsilon) = O\left( \frac{n^2 \log m}{m^2} \log \epsilon^{-1} \right).
	\end{align}
\end{theorem}

\begin{proof}
Divide the grid into squares of side length $m$.  Initially, each square contains $m^2$ agents.  Let $a_{i}$ refer to the agent whose initial location is $i$ and let $s(a_{i})$ refer to the square containing $i$.  The mobility of agent $a_i$ covers $s(a_{i})$ and intersects the squares adjacent to it.  For each pair of agents we must route $D(i,j) = 1/n^2$ units of flow.  We will do this by routing flows in L-shaped paths, as shown in Figure \ref{local_example} and \ref{local_example2}.  Since $a_{i}$'s mobility intersects the squares adjacent to $s(a_{i})$, there is a nonzero probability that agent $a_{i}$ will communicate with an agent $a_{i'}$ whose square $s(a_{i'})$ is adjacent to $s(a_{i})$.

Assign the $1/(m^2 n^2)$ units of flow to each agent $a_{i'}$ whose initial location is in the square adjacent to $s(a_i)$.  There are $m^2$ such agents.  Each agent then routes $1/(m^4 n^2)$ units of flow to each agent $a_{i''}$ in the next square along the L-shaped path.  The flow is routed only along edges $(j,j')$ such that $s(a_j)$ and $s(a_j')$ are different.  Each left-to-right edge carries flow from the $O(\sqrt{n}/m)$ squares to the left of it.  These flows are routed to the $O(n/m^2)$ squares to the right and above it for a total of $O(n^{3/2}/m^3)$ pairs $(i,j)$ that are routed through each square.  Each square has $m^2$ agents so there are $O(n^{3/2}/m)$ flows carrying $1/(n^2 m^2)$ per flow, so the load on the edge is
	\begin{align}
	f(i,j) = O\left( \frac{1}{\sqrt{n} m^3}\right).
	\end{align}
The same bound holds for down-to-up edges.

To find the 	capacity of these edges, we calculate the probability that agents $i$ and $k$ in adjacent squares average with each other.  The probability is $1/n$ to select agent $i$ and the overlap in agent $i$ and $k$'s mobility area is $\Omega(m^2)$, so the chance $i$ and $k$ are adjacent after moving is $\Omega(1/m^2)$.  With high probability there will be no more than $O(\log m)$ nodes for $i$ to choose from, so the chance of selecting $k$ is at worst $\Omega(1/\log m)$.  Thus:
	\begin{align}
	C(i,k) = \Omega\left( \frac{1}{n^2 m^2 \log m} \right).
	\end{align}
The maximum length of any flow is $O(\sqrt{n} / m)$, so the Poincar\'{e} inequality gives
	\begin{align}
	\frac{1}{1 - \lambda_2(\bar{W})} = O\left( \frac{n^2 \log m}{m^2} \right).
	\end{align}
	
\end{proof}

\subsubsection{Adding mobile agents}

The question motivating this work is this : how much can agent mobility improve the convergence speed of gossip or consensus algorithms?  Put another way, how much mobility is needed to gain a certain factor improvement in the convergence?  A simple model for which we can answer this question is the following: consider $n$ static agents in the $\sqrt{n} \times \sqrt{n}$ torus together with $m$ mobile agents whose mobility $\mu_i$ is uniform on the torus.  We use our techniques from earlier sections below to show that the averaging time of gossip in this model is $\Theta(n^2/m \log \epsilon^{-1})$, which for $m = n^{\alpha}$ is $\Theta(n^{2-\alpha})$.  For example, adding $\sqrt{n}$ mobile nodes can speed convergence by a factor of $\sqrt{n}$.

\begin{theorem}
Let the set of locations be given by the $\sqrt{n} \times \sqrt{n}$ discrete lattice on the torus $\mc{G} = (\mc{V},\mc{E})$.  Let there be $n+m$ agents $\mc{A} = \mc{S} \cup \mc{M}$ where the $n$ static agents $\mc{S}$ are positioned on the $n$ nodes of the torus and do not move. and the $m$ mobile agents $\mc{M}$ have mobility that is uniform on $\mc{V}$, where $m < n$.  Then the averaging time is given by
	\begin{align}
	\tave(n,\epsilon) = \Theta\left( \frac{n^2}{m} \log \epsilon^{-1} \right).
	\end{align}
\end{theorem}

\begin{proof}
We first show that for $i \in \mc{S}$ and $j \in \mc{M}$, the probability $P_{ij}$ that agent $i$ contacts agent $j$ and averages is $\Theta(1/n(m+n))$.  Agent $i$ is selected with probability $1/(m+n)$ and agent $j$ is in the neighborhood of agent $i$ with probability $5/n$.  Therefore:
	\begin{align}
	P_{ij} = \frac{5}{n (m+n)} \sum_{l=0}^{m-1} \frac{1}{5 + l} \prob(L = l),
	\end{align}
where $L$ is the the number of agents in $\mc{M}$ that land in the neighborhood of $i$.  The summation is just 
	\begin{align}
	\sum_{l=0}^{m-1} \frac{1}{5 + l} \prob(L = l) = \expe[1/(5+L)],
	\end{align}
which is clearly upper bounded by $1$, so 
	\begin{align}
	P_{ij} = O\left(\frac{1}{n(m+n)}\right).	
	\end{align}
Since $1/(5+L)$ is convex, Jensen's inequality can be used to obtain a lower bound: 
	\begin{align}
	\expe[1/(5+L)] \ge 1/\expe[5 + L] = 1/(5 + 5m/n).
	\end{align}
Therefore $P_{ij} = \Omega(1/n(m+n))$.  By symmetry, we have the same bound on $P_{ji}$.

To get the lower bound, consider the function $G : \mc{S} \cup \mc{M} \to \mc{S} \cup \{M\}$ that is the identity on $\mc{S}$ and merges $\mc{M}$ into a single state $M$.  We can bound the transition probabilities of the new chain using (\ref{eq:induced_trans}):
	\begin{align}
	\hat{W}_{Mi} &= \frac{1}{ \sum_{j \in \mc{M}} \pi(j) } \sum_{j \in \mc{M}} \pi(j) \frac{P_{ij} + P_{ji}}{2} \nonumber \\
	&= \Theta\left( \frac{1}{n (m + n)} \right) \\
	\hat{W}_{iM} &= \frac{1}{\pi(i)} \sum_{j \in \mc{M}} \pi(i) \frac{P_{ij} + P_{ji}}{2} %
	\nonumber \\
	&= \Theta\left( \frac{m}{n(m + n)} \right).
	\end{align}
For $i,k \in \mc{S}$ we have $\hat{W}_{ik} = \bar{W}_{ik}$.

The new chain is a torus plus an additional central node $M$.  The probability of transitioning from the torus to the central node is $\Theta( (m/n)/(m+n) )$ and for transitioning back it is $\Theta( (1/n)/(m+n) )$.  It can be seen (see the Appendix) that the relaxation time for this chain is $\Omega( n^2/m )$ via the extremal characterization in (\ref{eq:trel_extremal}).  Thus	%
	$\tave(n,\epsilon) = \Omega \left( \frac{n^2}{m} \log \epsilon^{-1} \right).$
We now turn to the upper bound.  As before, we construct a flow on the chain.  The demand between any two agents $(i,j)$ is $1/(n+m)^2$.  Since $P_{ij} = \Theta( 1/n(n+m) )$, the capacity 
\[
C(e) = \Theta( 1/n (n+m)^2),
\]
 for $e = (i,j)$.
We must now construct a flow that will yield an upper bound on the relaxation time of $n^2/m$.  For a pair of states $i \in \mc{S}$ and $j \in \mc{M}$ we assign $1/(n+m)^2$ to the direct path $(i,j)$.  For a pair $i \in \mc{S}$ and $j \in \mc{S}$ we split $1/(n+m)^2$ equally into the $m$ paths $(i,k,j)$ for $k \in \mc{M}$.    Finally, for $i \in \mc{M}$ and $j \in \mc{M} \cup \mc{S}$ we again route $1/(n+m)^2$ directly on $(i,j)$.  Then
	\begin{align}
	f((i,j)) = \left\{
	\begin{array}{ll}
	\frac{1}{(m+n)^2}  &  i,j \in \mc{M} \\
	0 & i,j \in \mc{S} \\
	\frac{1}{(m+n)^2} + \frac{n}{m} \frac{1}{(m+n)^2} &  i \in \mc{S},\ j \in \mc{S} \cup \mc{M} 
	\end{array}
	\right. \nonumber
	\end{align}
Therefore $\rho(F) = \Theta(n^2/m)$.  Since all paths are $\Theta(1)$, the Poincar\'{e} inequality implies that $\trel(\bar{W}) = O\left( n^2/m \right)$, so Theorem \ref{thm:boydthm} gives
	$\tave(n,\epsilon) = O\left( \frac{n^2}{m} \log \epsilon^{-1} \right)$.
\end{proof}

\subsection{Random geometric graphs}

\subsubsection{Bidirectional mobility}

We now turn to the case where some agents move horizontally and some vertically.  We will prove our results for the random geometric graph model, where $n$ nodes are initially placed uniformly in the unit square $\mc{G}$.  In the bidirectional  mobility model, before the gossip algorithm starts, each node flips a fair coin, is assigned to move horizontally or vertically, and moves like this throughout the process.  Note that this is a one-dimensional mobility model since each node is moving only horizontally or vertically throughout the execution of the gossip algorithm, never changing direction. Our result is that this mobility model is as good as complete node connectivity.

\begin{figure}[t]
\begin{center}
 \includegraphics[width=8cm]{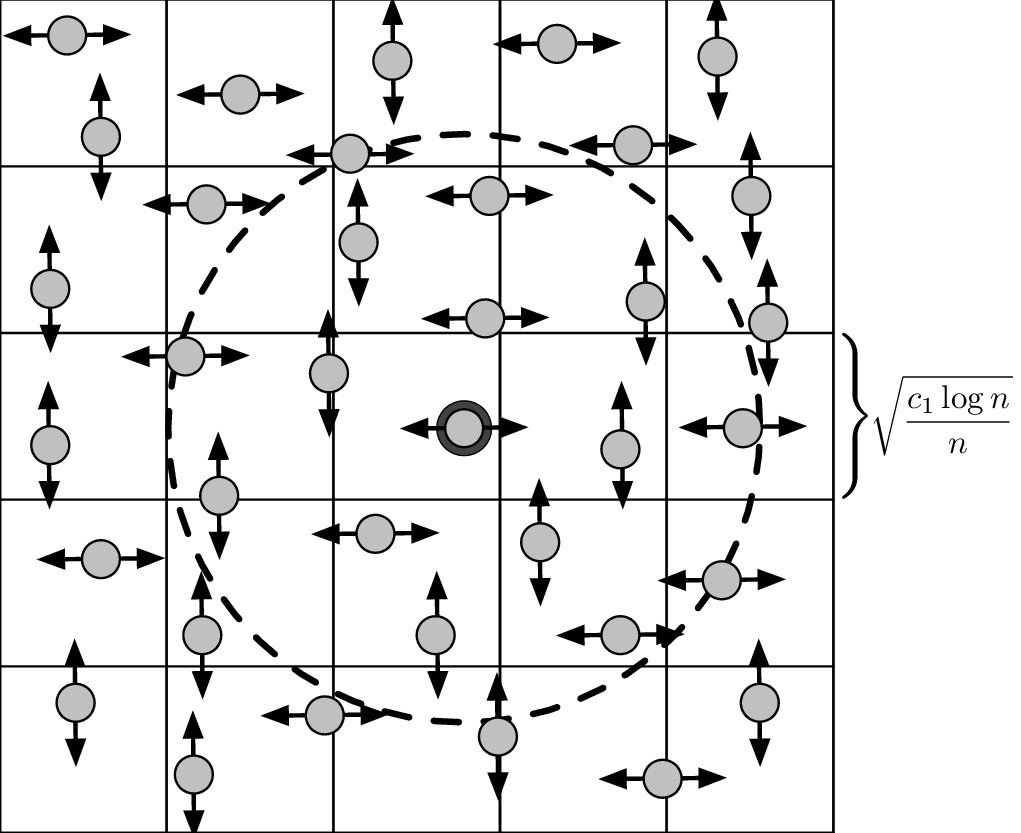}
  \caption{Random geometric graph example with bidirectional 1D mobility.}
  \label{rgg_example}
  \end{center}
\end{figure}

\begin{theorem}
Consider the gossip algorithm with $n$ agents under the random geometric graph model and bidirectional mobility.  We can choose a connectivity radius $r(n)=\Theta\left(\sqrt{\frac{\log n}{n}}\right)$ such that the the gossip averaging time is 
\begin{align}
\tave(n,\epsilon)=\Theta(n \log \epsilon^{-1}).
\end{align}
\end{theorem}

\begin{proof}
We start by partitioning the space into a grid of squares of size $c_1 \frac{\log n}{n}$.  Let $B_i$ denote the number of agents whose initial position was in square $i$. 

It is well known~\cite{Penrose:03rgg,GuptaK:00networks,ElGamalMPS:04tradeoff,BDTVPath,Avin:2007:CTM:1244475.1244725} that a combination of a Chernoff and a union bound, yields uniform bounds on the maximum and minimum occupancy of all the squares:
	\begin{align}
	\prob\left( \frac{c_1}{2} \log n \leq B_i \leq 2 c_1 \log n \ \forall i\right)
	\geq 
	1- n^{1-c_1/8} \frac{2}{c_1 \log n}. \nonumber
	\end{align}
By selecting $c_1 \geq 10$ we can show that all the squares have $\Theta(\log n)$ agents with probability at least $1-\frac{1}{n^2 \log n}$ so square occupancies are balanced even if the experiment is repeated $n^2$ times. We set the transmission radius to $r(n)= \sqrt{ 5 c_1 \frac{\log n}{n}}$ to guarantee that a agent in a square can always communicate with any agent in the four adjacent squares. 

Recall that initially each agent is assigned to be a horizontally moving or vertically moving node by flipping a coin and keeps this directionality throughout the process. Denote by $H_i$ the set of nodes that move horizontally and whose initial position was in the $i$-th row of squares.  These agents always stay in the $i$-th row.  Similarly, let $V_i$ be the set of agents who move vertically in the $i$-th column of squares.

Each square contains in expectation $c_1 \log n$ nodes and there are $\sqrt{\frac{n}{c_1 \log n}}$ squares in each row and column. Since each node flips a fair coin and is assigned in a vertically or horizontally moving class, the expected cardinalities will be:
	\begin{align} 
	\stexp |H_i|
	= \stexp|V_i|
	= \frac{1}{2}c_1 \log n \sqrt{\frac{n}{c_1 \log n}}
	= \Theta( \sqrt{n \log n}).
	\label{eq:numHVguys}
	\end{align} 
Using standard Chernoff bounds we can show that the cardinalities of $|H_i|,|V_i|$ 
are sharply concentrated near their expectation. 

Theorem \ref{thm:boydthm} shows that the averaging time of the gossip algorithm is bounded by the inverse spectral gap (relaxation time) of the average matrix $\bar{W}$, where the expected matrix $\bar{W}=\stexp{W(s)}$ is computed over mobility of the nodes and random selection of which nodes are gossiping. 

We now proceed to bound the spectral gap using a canonical flow and we need 
to select paths for every pair of states for the Markov chain defined by $\bar{W}$.  The state space is the set of $n$ agents and $\pi(i)=1/n$ for each agent $i$ since $\bar{W}$ is doubly stochastic. The \emph{capacities} of the edges will be proportional to the entries of $\bar{W}$ (see \ref{eq:W_entries}), where $\bar{W}_{ij}$ is the average of the probabilities $P_{ij}$ and $P_{ji}$, measuring how often agents $i$ and $j$ are pairwise
averaged. For each pair of agents $(i,j)$ we must specify how to satisfy the demand $D(i,j) = n^{-2}$ by assigning flows to some (appropriately chosen) paths in $\mc{P}_{ij}$. 

Our flow construction uses four different cases depending on whether $i$ and $j$ move horizontally or vertically:

\begin{figure}[t]
\begin{center}
 \includegraphics[width=8cm]{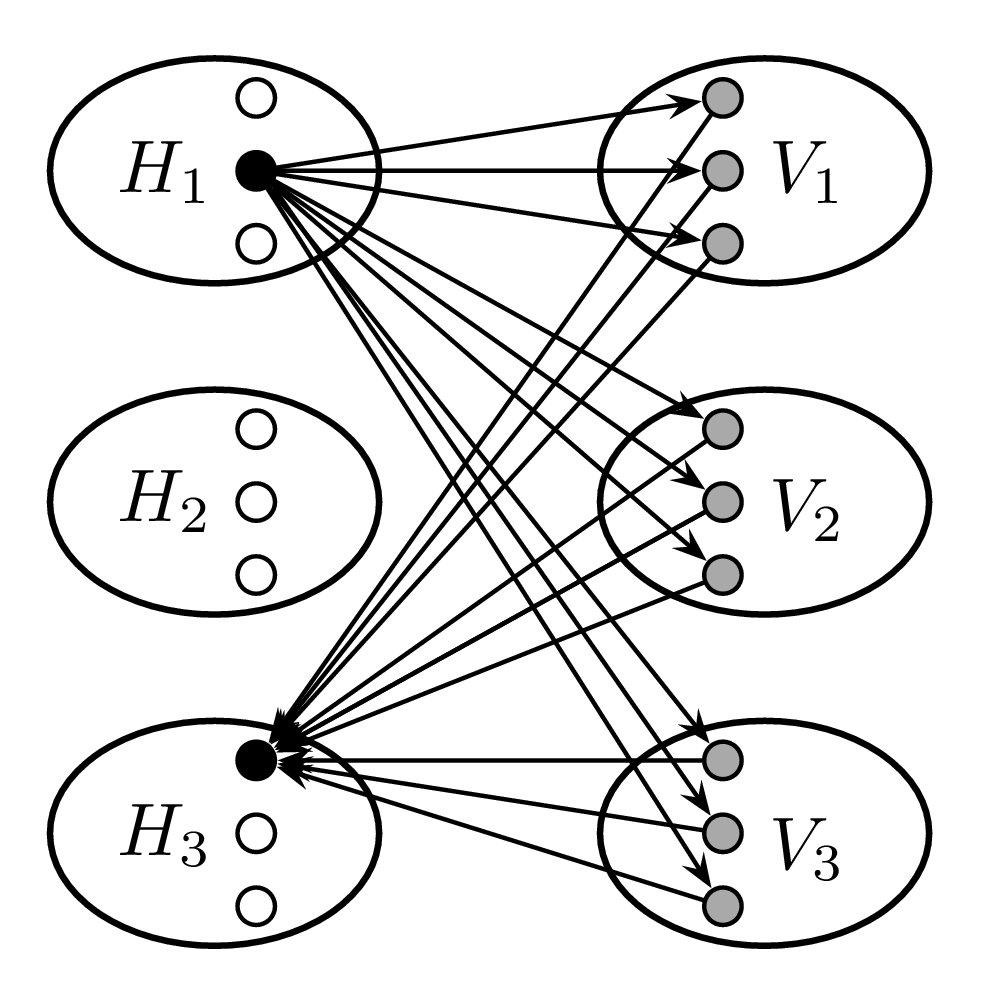}
  \caption{Routing flow from a node in $H_1$ to a node in set $H_3$.  The flow is routed from the node in $H_1$ to all nodes in the sets $\{V_i\}$ and then back to the node in $H_3$.}
  \label{fig:fg_flow_HV}
  \end{center}
\end{figure}

\paragraph{Case 1}  Suppose $i \in H_k$ and $j \in H_l$.  To satisfy the demand $n^{-2}$ node $i$ assigns $\Theta(n^{-3})$ units to each path $(i, v, j)$, where $v \in V_r$ for some $r$.  There are $\Theta(n)$ agents who move vertically, so the total flow that reaches $j$ can be made equal to $n^{-2}$.  See Figure \ref{fig:fg_flow_HV}.

\paragraph{Case 2}  Suppose $i \in V_k$ and $j \in V_l$.  This is the same as the previous case, except that $\Theta(n^{-3})$ units are assigned to each path $(i,h,j)$ for $h \in H_r$.

\paragraph{Case 3}  Suppose $i \in H_k$ and $j \in V_l$.  To satisfy the demand $n^{-2}$ assign $n^{-2}$ to the direct path $(i,j)$.  

\paragraph{Case 4}  Suppose $i \in V_k$ and $j \in H_l$.  We again assign $n^{-2}$ to the direct path $(i,j)$.

\noindent Our construction therefore only uses the edges in the graph between $H$ sets and $V$ sets. In other words \emph{it is only the averaging between nodes that move vertically with nodes that move horizontally} that allows information to spread fast in the network. The averaging between two nodes in $H$ or $V$ could be omitted and still the bound would not change in order. 
  The total load on an edge $e = (h,v)$ between a horizontal moving agent and a vertical moving agent is the sum of the direct flow $(h,v)$, the the sum of the flows $(h,v,j)$ for all horizontal moving $i$ and $(i,h,v)$ for all vertical moving $i$.
	\begin{align}
	f(e) &= \frac{1}{n^2} 
	+ \Theta\left(\frac{1}{n^3}\right) \sum |V_r|
	+ \Theta\left(\frac{1}{n^3}\right) \sum |H_r| \nonumber \\
	&= \Theta\left( \frac{1}{n^2} \right).
	\end{align}
The same bound holds for $e = (v,h)$.

Finally, we calculate the capacity for the edges $(v,h)$.  It is sufficient to calculate a lower bound on the probability that agents $v \in V_k$ and $h \in H_l$ average.  Agent $v$ is selected with probability $1/n$.  Based on our assumptions on the communication radius, $v$ can communicate with $\Theta(\log n)$ neighbors.  The probability that $v$ lands in a row within $r(n)$ of row $l$ is $\Theta(\sqrt{n^{-1} \log n})$ and the probability that $h$ lands within $r(n)$ of row $k$ is also $\Theta(\sqrt{n^{-1} \log n})$.  Therefore we have
	\begin{align}
	P_{vh} = \Omega\left( \frac{1}{n} \sqrt{\frac{\log n}{n}}
	\sqrt{\frac{\log n}{n}}
	\frac{1}{\log n}
	\right) = \Omega\left( \frac{1}{n^2} \right).
	\end{align}
The capacity of each edge $(v,h)$ is then $C(e) = \Omega(n^{-3})$.  By symmetry, the same formulae hold for $(h,v)$.

We can now calculate the overload for this flow on any edge $e = (v,h)$:
	\begin{align}
	\frac{ f(e) }{ C(e) } = \frac{\Theta(n^{-2})}{\Omega(n^{-3})} = O\left( n \right).
	\end{align}
Since this holds for all edges we have $\rho(F) = O(n)$.  The maximum length of any path used in the flow is $2$, so by the Poincar\'{e} inequality we have
	\begin{align}
	\trel(\bar{W})
	= \frac{1}{1-\lambda_2(\bar{W})}
	= \rho(F) l(F)
	= O(n).
	\end{align}
Theorem \ref{thm:boydthm} gives the result.
\end{proof}

One intuition for this result is that bidirectional mobility enables the construction of ``short'' routes between all pairs of agents.   We can derive the identical result for the torus using the same arguments.  Under bidirectional mobility the averaging time for the torus is $O(n \log \epsilon^{-1})$, which is the same as full mobility.

\subsubsection{Unidirectional mobility}

We now show that unidirectional mobility does not improve the scaling performance for random geometric graphs.  This is proved in the same way as the analogous result for the torus.

\begin{corollary}[Random geometric graph with 1D mobility]
Consider gossip on the random geometric graph with $n$ agents with the 1D unidirectional mobility model.  Then for this gossip algorithm,
	\begin{align}
	\tave( n, \epsilon ) 
		= 
		\Omega \left( \frac{n^2 \log \epsilon^{-1}}{\log n} \right).
	\end{align}
\end{corollary}

\begin{proof}
We first divide the unit square into sub-squares of side length $c_1 \sqrt{\frac{\log n}{n}}$ for some constant $c_1$.  This creates a $\Theta\left(\sqrt{\frac{n}{\log n}}\right) \times \Theta\left(\sqrt{\frac{n}{\log n}}\right)$ torus on which the mobility can be defined.  We must first characterize the Markov chain corresponding to the gossip algorithm under the 1D unidirectional mobility model.  If we set the communication radius to $c_2 \sqrt{\frac{\log n}{n}}$ then an agent in the $i$-th row of sub-squares can communicate with agents in rows $\{i-c_3, \ldots, i+c_3\}$, where $c_3$ is again a constant.  Moreover, each sub-square will have $\Theta(\log n)$ agents with high probability.  Therefore we can upper bound the probability that an agent in row $i$ will average with an agent in one of the rows $\{i-c_3, \ldots, i-1, i+1, \ldots, i+c_3\}$:
	\begin{align}
	\beta_{ij} = O\left( \frac{1}{n} \times \sqrt{\frac{\log n}{n}} \times \frac{1}{\log n} \right).
	\end{align}
Thus the chance a given agent averages with someone not in their row of sub-squares is $O(1/\sqrt{n^3 \log n})$.  

As in the torus, we apply the induced chain method using the partition that merges each row of sub-squares.  This creates a new Markov chain with $\sqrt{n/\log n}$ states that is a kind of cycle where there are positive transition probabilities from state $k$ (corresponding to the $k$-th row) to states $l \in \{i-c_3, \ldots, i+c_3\}$.  From the analysis of the torus we can see that from row $k$ to $l$:
	\begin{align}
	\hat{W}_{kl} &= \frac{1}{ \sum_{i : F(i) = k} \pi)(i) } \sum_{i : F(i) = k} \sum_{j : F(j) = l} \pi(i) W_{ij} \\
	&= \sqrt{ \frac{n}{\log n}} \cdot n \log n 
		\cdot \frac{1}{n} \cdot
		O\left( \frac{1}{n^{3/2} \sqrt{\log n} } \right) \\
	&= O\left( \frac{1}{n} \right).
	\end{align}
Let $\beta$ denote this transition probability.  The matrix of this new chain is still circulant and generated by the vector
	\begin{align}
	(\beta, \ldots, \beta, 1 - 2 c_3 \beta, \beta, \ldots, \beta, 0 \ldots 0).
	\end{align}
The DFT and Taylor expansion again gives the bound on the second-largest eigenvalue:
	\begin{align}
	\lambda_2(\hat{W}) &= 1 - \beta \cdot O\left( \frac{\log n}{n} \right) \\
		&= 1 - O\left( \frac{\log n}{n^2} \right).
	\end{align}
Therefore $\trel(\hat{W}) = \Omega( n^2/\log n )$.
\end{proof}

\section{Experiments and simulations}

We can gain some intuition about the benefits of mobility via simulations.  All simulations are for a torus with a linearly varying field.  Our first main result was a lower bound that shows horizontal mobility is as bad as no mobility in terms of convergence.  This is illustrated in Figure \ref{fig:torusNoH}, where we can see that for a range of network sizes the error under horizontal mobility is close to that of the torus with no mobility.  Indeed, as the network size gets larger, the gap vanishes, which suggests that our analysis is tight for this example.  Our second major result was a positive one; the bidirectional mobility model was nearly as good as full mobility.  This is illustrated in Figure \ref{fig:torusFullHV}.  Although there is a gap between the error decay under the two mobility models, for a fixed error the number of iterations needed to achieve that error is at most a constant factor more for the bidirectional mobility model.%

\begin{figure}
\begin{center}
\includegraphics[width=3.0in]{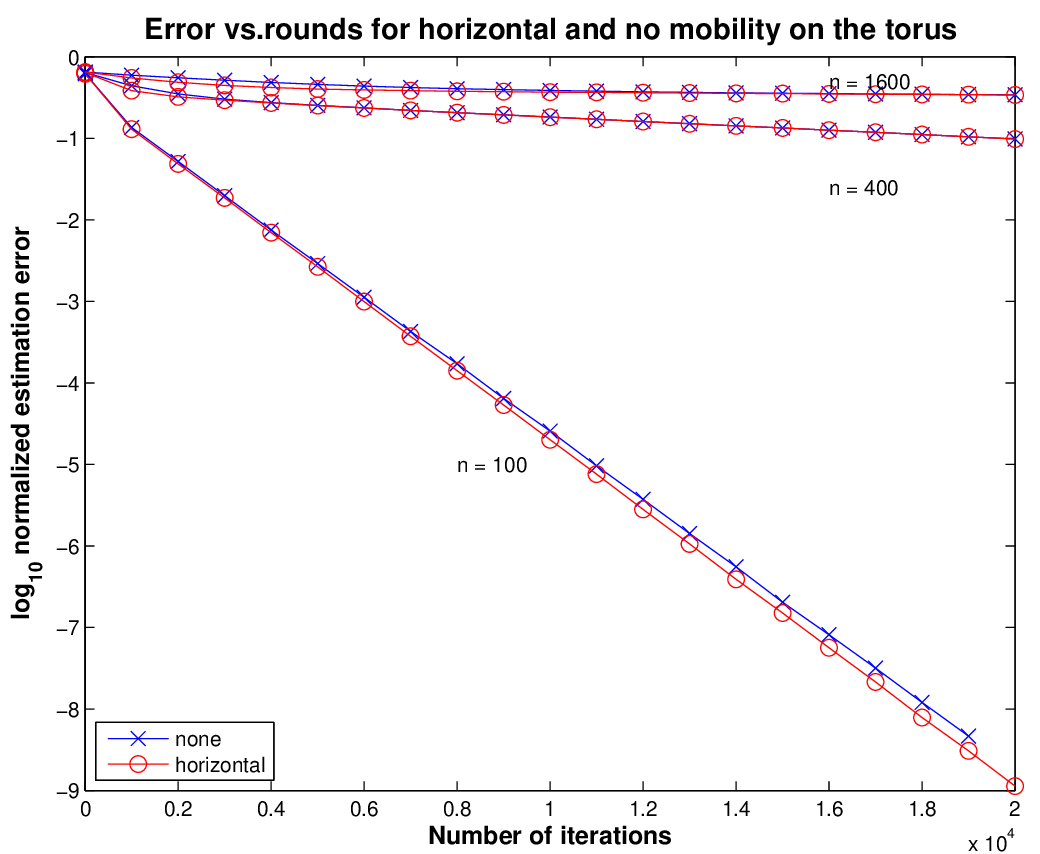}
\caption{ Log average error versus number of iterations of the gossip algorithm for the torus with no mobility and with horizontal mobility.  As the graph size increases, the gap between the two algorithms vanishes. \label{fig:torusNoH}}
\end{center} 
\end{figure}

\begin{figure}
\begin{center}
\includegraphics[width=3in]{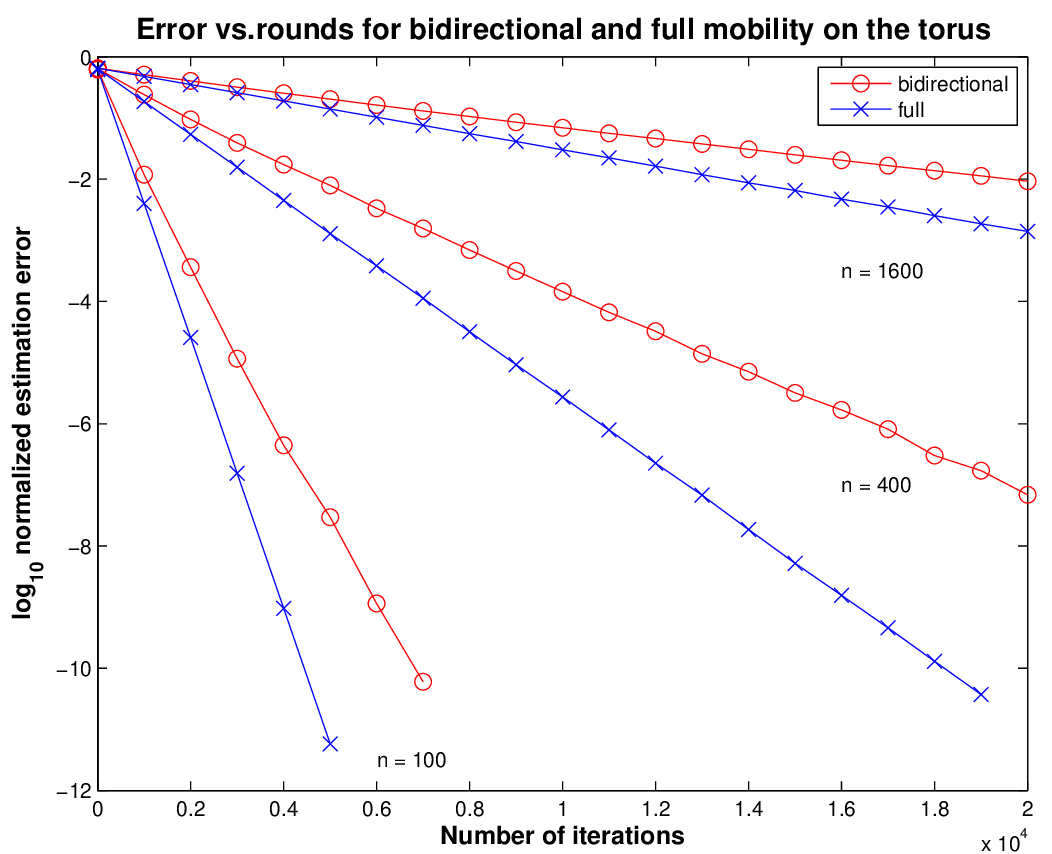}
\caption{ Log average error versus number of iterations of the gossip algorithm for the torus with full mobility and with bidirectional mobility.  As the graph size increases, the gap between the two algorithms shrinks. \label{fig:torusFullHV} }
\end{center}
\end{figure}

Our final result was that adding $m$ mobile agents to a static grid with $n$ agents gives a convergence time of $\Theta(n^2/m \log \epsilon^{-1})$.  Figure \ref{fig:addguys} shows how adding only a few additional mobile agents can dramatically improve the speed of convergence.  As we add more nodes, $\log \epsilon$ decreases linearly, which corresponds to an exponential decay in the average error.  This suggests that even in large networks, investing in a small number of mobile agents can yield a major benefit in convergence time.

\begin{figure}
\begin{center}
\includegraphics[width=3in]{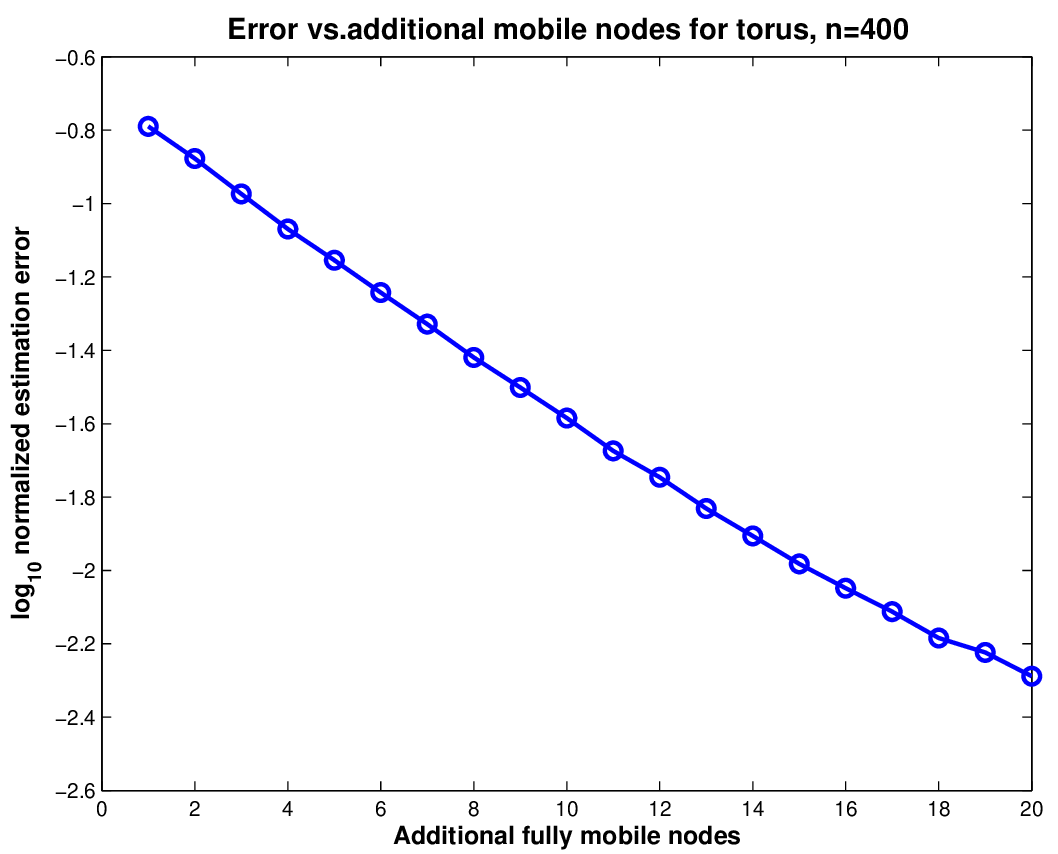}
\caption{ Adding a few mobile nodes to a static grid can exponentially decrease the estimation error for a fixed number of iterations (20000). \label{fig:addguys} }
\end{center}
\end{figure}

The examples that we consider in this paper are simplifications of real network topologies and real mobility models.  It is important to understand how unrealistic these models are.  We simulated the difference between the lattice on the torus versus a $\sqrt{n} \times \sqrt{n}$ grid.   Figure \ref{fig:torusgrid} shows the error after a fixed number of iterations for increasing grid sizes.  Although the algorithm converges faster on the torus, the gap decreases with larger network size.  A second question is how the \textit{random walk} mobility model \cite{GrossglauserV:06ease,latticeRW} relates to the mobility model in this paper.  In order to analyze gossip under such a mobility model, we would need to prove new convergence result for the iterated random matrix products that characterize the evolution of the agents' estimates.  It is clear that if each agent moves according to a random walk and the number of steps taken between each gossip iteration is longer than the mixing time of the random walk, then random walk mobility is equivalent to the mobility models considered here.  However, for a smaller number of steps, the simulations of the speed of convergence of the algorithm are inconclusive, as there appears to be a dependence on the initial configuration of agents' values.  We leave as an open question how to bound the performance Markov random walk models.

\begin{figure}
\begin{center}
\includegraphics[width=3in]{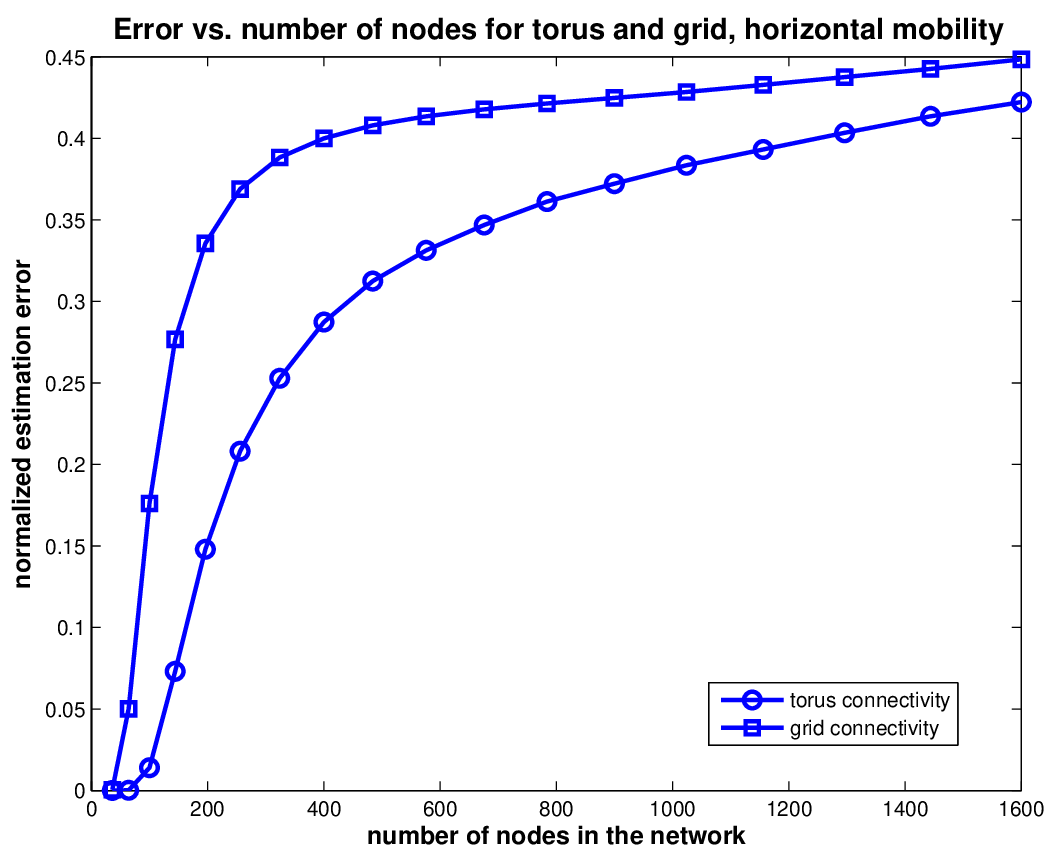}
\caption{ Gap between the torus and grid versus grid size after 5000 pairwise iterations, averaged over 100 trials, using the uniform horizontal mobility model. \label{fig:torusgrid} }
\end{center}
\end{figure}

\section{Discussion and future directions}

In this work we investigated how agent mobility impacts the convergence speed of distributed averaging algorithms by developing new analytical tools derived from the theory of Markov chains.  Using these tools we could show that different mobility patterns can have dramatically different effects depending on the overlap of the mobility paths.  Perhaps surprisingly, even a sublinear number of mobile nodes can change the order of gossip messages required for convergence.  We note that ``mobility'' in our model is a variety of time-varying network topology that in practical implementations need not come from the physical mobility of the agents, but can be induced by structured variations in the topology.

The class of mobility models which are amenable to our analysis makes a strong assumption on the speed of the mobility or delay-tolerance of the gossip algorithm.   One interesting direction for future research involves understanding the benefits of mobility for more realistic mobility models. It is possible that general mobility models with memory 
are tractable to analysis if the mobility is driven by a Markov chain since this would 
integrate naturally with the Markov structure of the averaging process.  Proving that these systems reach a consensus could follow from more general results about the corresponding stochastic process \cite{TouriN11:flow}.  
 We conjecture that random walk models with slower mixing times will yield smaller benefits, and that our independent (fast mixing) model is an upper bound.  For these models, modifying the pairwise gossip paradigm 
(c.f. \cite{BDTVPath}) may yield a greater benefit then relying on mobility alone.  The impact of node mobility on distributed optimization and general message-passing algorithms on probabilistic graphical models would also be a very interesting research direction.  

Another interesting direction is understanding the impact of mobility for more general message-passing algorithms such as distributed convex optimization. The analysis of \cite{Moallemi07} obtains a convergence theorem similar to the spectral gap and it would be interesting to investigate the scaling behavior of the number of required iterations for the min-sum algorithm to optimize a convex function under out node mobility models.

\appendix

We will construct a $g$ in (\ref{eq:trel_extremal}) to show that the mixing time of a torus plus an additional central node $M$ with transition probabilities $\Theta( (m/n)/(m+n) )$ to $M$ and $\Theta( (1/n)/(m+n) )$ away from $M$ along with transitions $\Theta(1/n)$ between neighbors in the torus has relaxation time $\Omega(n^2/m)$, where $m < n$.  The stationary distribution for this chain has probability $\pi(i) = \Theta(1/(m+n))$ on the nodes $i = 1, 2, \ldots, n$ of the torus and $\pi(M) = \Theta(m/(m+n))$ on $M$.  Let $g(M) = 0$ and $g$ be constant on each column of the torus with the values on the columns being $\{-\alpha, -\alpha+1, \ldots 0, 1, 2, \ldots, \alpha, \alpha, \alpha-1, \ldots, -\alpha+1, -\alpha \}$ for $\sqrt{n}$ even and $\{-\alpha, -\alpha+1, \ldots, \alpha, 0, \alpha, \alpha - 1,\ldots, -\alpha\}$ for $\sqrt{n}$ odd, where $\alpha = \Theta(\sqrt{n})$.  Then clearly $\sum \pi(i) g(i) = 0$.  We can calculate the numerator and denominator in (\ref{eq:trel_extremal}):
	\begin{align}
	\sum_{k} \pi(k) g(k)^2 &= \frac{1}{m+n} \sqrt{n} 4 \sum_{i=0}^{\alpha} i^2 \nonumber \\
	&= \Theta\left(\frac{n^2}{m+n}\right) \\
	\mc{D}(g,g) &= \frac{m/n}{(m+n)^2} \sqrt{n} 4 \sum_{i=1}^{\alpha} i^2 \nonumber \\
	&\qquad \qquad + \frac{1/n}{m+n} \sqrt{n} 4 \sum_{i=1}^{\alpha} 1 \nonumber \\
	&= \Theta\left( \frac{1 + \frac{mn}{m+n}}{ m+n } \right).
	\end{align}
Dividing gives the result.

\bibliographystyle{IEEEtran}

\bibliography{mob_gos_abrv}

\end{document}